\documentclass[12pt]{article}
\input{epsf.tex}

\begin{document}

\begin{center}
{\bf PRACTICAL LIMITS ON ACHIEVABLE PRECISION OF SOME
NUCLEAR-PHYSICS PARAMETERS DETERMINATION
} \end{center}
\begin{center}
{\bf  A.M. Sukhovoj, V.A. Khitrov}\\
\end{center}\begin{center}
{\it  Joint Institute
for Nuclear Research, 141980, Dubna, Russia}\\
\end{center}
\begin{abstract}
The status of experiments on determination of level density and
partial widths of the nuclear reaction products emission in
diapason of nucleon binding energy is presented.
There are analyzed the sources and magnitude of probable
systematical uncertainties of their determination.
The maximally achievable precision of these parameters is
estimated, as well.

There is considered ability of new method for determination of
distribution parameters of neutron resonances
reduced widths in order to distinguish their groups with the
same structure of wave functions.
It was obtained in both cases that the insufficient value of
maximally achievable precision of the parameters of the
experimental data analysis does not allow one to obtain reliable
and detailed information on the studied nuclear properties
-- its entropy, strength functions of nuclear products emission
and dominant level structure above $\approx 0.5B_n$.
 
\end{abstract}

\section{Introduction}\hspace*{16pt}

The problem of determination of level density $\rho$ and partial
widths $\Gamma$ of gamma-quanta or nucleons interaction with
excited nucleus is an object of numerous experiments.
One can say the same and about determination of nuclear properties
from parameters of neutron resonances of any stable or long-living
target-nuclei.

In the first case, the special problem and importance is a
necessity of their determination in the excitation energy region
where resolution of existing spectrometers is insufficient for
distinguishing of individual nuclear levels.
Accordingly, only the spectra of reaction products $S$ and their
cross sections $\sigma$ are available for experimental
determination.
Their observables are determined by product $\rho \Gamma$,
entering as a parameter in corresponding functionals -- the total
gamma-spectra, averaged cross sections, spectra of two-step
reactions.
The $\rho$ and $\Gamma$ values can be extracted only from these
data by solution of reverse task of mathematical analysis.
If one does not consider the problem of precision of the model
set functional relations $S=F(\rho, \Gamma)$ (or $\sigma=F(\rho, \Gamma)$), then
unambiguous (asymptotically precise) determination of $\rho$ and
$\Gamma$ is possible only under the condition that the matrix
$L_m=J^{t} W J$ is not degenerated.
Here is postulated that the Jacobi matrix $J$ contains all the
possible information on the case under consideration and the
weight matrix $W$ - diagonal. 

If the number of real parameters in system of equations
$S=F(\rho, \Gamma)$ (or $\sigma=F(\rho, \Gamma)$)
is more than the number of experimental
points, $L_m$ is always degenerated.
In practice, there were not found any systems of non-degenerated
non-linear equations connecting $\rho$, $\Gamma$ with the functions
under study.

However, because of nonlinearity of function $F$, the region of
the $\rho$ and $\Gamma$ values can be final even in this case.
Minimization of uncertainty of the sought parameters in this case
can be achieved only by choice of the most effective experiment.
This conclusion follows from analysis of mathematical relations
between the parameters and measured functions in one- and two-step
experiments \cite{Mo96}.
Estimation of the possible achievable precision in determination
of both $\rho$ and $\Gamma$ can be obtained only at analysis of
concrete possibilities of existing methods for determination of
these parameters.

\section{Partial widths and level density}\hspace*{16pt}

1. The spectra of evaporation nucleons in unresolved energy region
contain bigger number of unknown parameters than corresponding
to them experimental points.
By this, maximal width of the interval of possible values of the
sought parameter $\rho$ cannot be determined (or even limited)
without involving of additional information.
Practically in all performed within this method experiments,
the not measured width $\Gamma$ of nucleon emission at transition
of a nucleus from level $U_i$ to level $U_f>0$ was changed by
the calculated value of this parameter for $U_f=0$.

Real precision of such subjective notions on the $\Gamma$ values set
in this way is unknown. Therefore, all accumulated from
corresponding analysis information contains unknown uncertainty.
Its estimation - 15\% (published in \cite{Vona83}) can be
related only to the widths observed in the spectra of evaporation
nucleons as resolved peaks.
This conclusion unambiguously follows from analysis of form of
dependence of differential cross section of emission of
evaporation nucleons with energy $E_N$ by nucleus with excitation
energy $U>E_N$.

In  Hauser-Feshbach notion, the considered cross section is
determined by sum over initial and final levels of products like
$\Gamma_b(U,J,\pi,E,I,\pi)\rho_b(E,I,\pi)$  for final reaction
product $b$ \cite{Vona83,PRC76}:

\begin{eqnarray}
{\frac{d\sigma}{d\varepsilon_b}(\varepsilon_a,\varepsilon_b)=}
\sum_{J\pi}\sigma^{\mathrm{CN}}(\varepsilon_a)\,\frac{\sum_{I\pi}
\Gamma_b(U,J,\pi,E,I,\pi)\rho_b(E,I,\pi)}{\Gamma(U,J,\pi)}
\end{eqnarray}
where
\begin{eqnarray}
\lefteqn{\Gamma(U,J,\pi)=\sum_{b^\prime}\left(\sum_k \Gamma_{b^\prime}(U,J,\pi,E_k,I_k,\pi_k)\,+\right.}\\
&&\left.\sum_{I^\prime\pi^\prime}\int_{E_c}^{U-B_{b^\prime}}dE^\prime\,
\Gamma_{b^\prime}(U,J,\pi,E^\prime,I^\prime,\pi^\prime)\,
\rho_{b^\prime}(E^\prime,I^\prime,\pi^\prime)\right).\nonumber
\end{eqnarray}

It follows from (1) and (2), that at presence of completely unknown
systematical error $\delta$ of the calculated value of the width $\Gamma_b$,
the experimental
cross-section $\frac{d\sigma}{d\varepsilon}$ can be precisely
reproduced only by use of the level density with adequate
systematical error. 
If the calculated $\Gamma_{cal}$ and unknown experimental
$\Gamma_{exp}$ widths are connected by the relation
$\Gamma_{cal}=\Gamma_{exp}(1+\delta)$, then the mistaken level
density $\rho_{cal}=\rho_{exp}/(1+\delta)$ is to be used in
calculation for precise reproduction of cross section.
Of course, the unknown relative error $\delta$ of the calculated
width depends on excitation energy of final nucleus and can depend
on spin of levels which are connected with each other by emission
of reaction product.
In this case, the error $\delta$ is the weight average.
If one determines it by means of relation:
$(1+\delta)=\rho_{es}/\rho_{2\gamma}$ which connects level density
$\rho_{es}$ from evaporation spectra \cite{Vona83,PRC76} with
density $\rho_{2\gamma}$ determined from cascade intensity
\cite{Meth1,PEPAN-2005}, then it is possibly to estimate directly
the error of calculated cross section $\frac{d\sigma}{d\varepsilon}$
at any excitation energy of final nucleus.
The corresponding data were obtained for two nuclei and presented
in \cite{W181,Ni60}.

2. By now, the Norway collaboration has measured the total gamma
spectra of depopulation of the levels in wide excitation energy 
interval. In correspondence with the method described in \cite{NIM},
the authors of the following experiments extract from them the
spectra corresponding to the first gamma-quantum of cascade.
By this, it is postulated that the strength function
$f=\Gamma/(D E^3_\gamma)$ for arbitrary partial gamma-width is
determined only by its multipolarity  and does not depend on
nuclear excitation energy.
The authors suppose without a proof that the use of the known
values of the total radiative width of neutron resonances,
mean spacing between them $D$ and density of low-lying levels
permits one to get unbiased estimation of $\rho$ and $\Gamma$.
But, according to \cite{Oslo-err}, the matrix $L_m$ in this
case is  also degenerated. And the existing \cite{PEPAN-2005}
experimental data on cascade population of large set of excited
levels point to dependence of  $k=f/A^{2/3}$ on structure of
decaying and excited by gamma-transitions levels.
Therefore, there is no possibility to obtain asymptotically
zero uncertainty within the framework of the existing method
\cite{NIM}. Besides, the authors did not estimate neither the
value of systematical error nor required precision of their
experiment. It is done by us only in \cite{TotSpe}.

3. Any two-step reaction, for example, cannot give asymptotically
zero uncertainty in determination of $\rho$ and $\Gamma$.
Nevertheless, even in presence the asymptotical uncertainty of
the derived from it data
provides obtaining of quite acceptable information on the
gamma-decay parameters.
In particular, the width of interval of the possible $\rho$ and
$\Gamma$ values can be equal to some tens percents for the
cascade of two gamma-quanta proceeding between neutron resonance
and a group low-lying levels at zero total error of determination of the
cascade intensity.
Systematical error related with dependence of the radiative
strength functions of the cascade dipole gamma-transitions on
structure of initial and final levels, in two-step reactions can
be taken into account, at least, partially \cite{PEPAN-2005}.

4. Hence, it is necessary to develop and realize  new independent
methods of the $\rho$ and $\Gamma$ experimental determination.
It is no sense to realize new experiment in
the one-step variant.
In practice, it is necessary to pass to registration of cascades
of not only two successively emitted gamma-quanta but also of
three and more \cite{2008KIZY}. The more general solution of this
problem is to analyze the spectra of the two-step reactions with
registration of nucleon products at the first step of nuclear
reaction \cite{W181}.

Serious problem in the experiments on determination of the most
probable $\rho$ and $\Gamma$ values from two-step gamma-cascades
is unknown influence of structure of the initial cascade level on
the primary gamma-transition width -- 
all corresponding data on $\rho$ and $\Gamma$ were obtained from
analysis of intensity of cascades following thermal neutron
capture (only 1 or 2 initial levels are excited with visible
probability). As the most probable explanation, just the structure
of wave function of neutron resonance is the cause of considerable
variations of the strength function parameters in nuclei with
different mass \cite{appr-k}.
Dependence of strength functions on structure of intermediate
cascade level in wide region of its energy unambiguously follows
from the data \cite{PEPAN-2005}. It follows also a necessity to
reveal a degree of dependence of strength functions and on
structure of initial level.
Id est, the results presented in \cite{PEPAN-2006, YAF73} require
one to perform analysis of parameters of neutron resonances in
order to discover the dependence of this type and its possible
influence on dynamics of nuclear reaction. The total intensity of
all the primary transitions, naturally, equals 100\% per decay.
Therefore, the expected effect can only change the form of energy
dependence of cascade intensity in different resonances.

Level density in small, as compared with $B_n$ excitation energy interval
$\Delta U=U-B_n$, can be presented as decomposition in the row:
\begin{equation}
\rho(U)=\rho(B_n)+\frac{d\rho}{dU} \Delta U+....
\end{equation}

In the ideal case, the analysis of the reduced neutron widths must give the 
$\rho(B_n)$ and $\frac{d\rho}{dU}$ values with minimally possible error.
The maximally precise $\rho(B_n)$ values are requred by normalization of
its functional dependence for $U <B_n$.
Reliably determined negative values of $\frac{d\rho}{dU}$ would testify to
undoubted change in structure of neutron resonances and above
$B_n$ also.

\section{Reduced neutron widths}\hspace*{16pt}

The main grounds of this problem are the following:

(a) break, at least, of the first 3-4 Cooper pairs occurs in any nucleus
discretely, with the interval by excitation energy being some
less 2$\Delta_0$
($\Delta_0=12.8/\sqrt{A}$) \cite{PEPAN-2006,YAF73};

(b) moreover, there is observed excessive variation of parameters
of the best approximation of radiative strength functions
\cite{appr-k}. Besides, the forms of energy dependence of the
two-step cascade intensities and total gamma-spectra following
thermal neutron capture change, possibly, cyclically and, possibly, not 
accidentally. 

1. Inevitable errors of experiment and random fluctuations of both
the primary gamma-transition intensities and obtained $\rho$ and
$\Gamma$ values do not allow one to connect break thresholds energy
$U_{th}$ of Cooper pairs and $B_n$ .
I. e., to fix the most probable structure of nucleus near $B_n$
(number of broken pairs, difference of neutron binding energy and
break threshold of the last pair and so on).

According to the results \cite{YAF73}, correlation between level
densities of vibration and quasi-particle types continuously changes
at change of excitation energy.
This conclusion follows from theoretical notions about form of
energy dependence of density of levels of quasi-particle and
vibration type and from the lowest $\chi^2$ value at approximation
of the Dubna data set on level density as compared with
approximations \cite{PEPAN-2006,Prep196}.
This circumstance must change strength function of the primary
gamma-transitions following decay of the excited levels above
$B_n$ because of change in components of wave function which
determine the value of its matrix element \cite{PEPAN-1972}.
But the theory like quasi-particle-phonon model of nucleus cannot
predict quantitatively details of this process now.

Therefore, one can accept determinate on the whole character of
change of wave function of neutron resonances (and other
high-lying levels) as a working hypothesis qualitatively
explaining enumerated above aspects of nuclear investigation.
This change can accordingly influence the distribution of the
reduced neutron widths $\Gamma_n^l$ for orbital momenta
$l=0, 1...$.
Only experiment can determine whether this hypothesis is true or
wrong.

2. According to conclusions \cite{YAF73}, structure of resonances
changes continuously (probably -- not monotonously).
That is why, potential dependence of $\Gamma_n^l$ on the neutron
(proton) resonance energy, at worst for investigation cases
is smooth and in some interval
of their energy -- weak.

In practice, it is impossible to obtain information on,
for example, the value of the most important components of
wave function of neutron resonance.
Therefore, any data on appeared problem can be derived only from
analysis of distribution parameters of reduced neutron widths
measured in small energy interval with different type errors and
distortions.

Id est, it is necessary to solve analogous to the previous case
problem distorted by its nature, namely:

(a) to minimize a degree of model dependence at analysis of
distributions $\Gamma_n^l$;

(b) to develop new method for analysis of available experimental
information and to determine all region of its possible solutions;

(c) to develop new methods of investigations.

Random character of the $\Gamma_n^l$ values observed in experiment
is grounded experimentally and theoretically.
The first part of the ground  -- the use of principles of
mathematical statistics and its criterions, the second --
development of model description of experimental data.
In particular  -- theoretical investigation of fragmentation
regularities of any nuclear state over higher-lying levels
\cite{MalSol}. 

As a result, the Porter-Thomas hypothesis \cite{PT} transformed
in immutable axiom that the random variation of reduced widths is
described by $\chi^2$- distribution with degree of freedom
$f \approx 1$.

For absolute correctness of hypothesis \cite{PT}, it is necessary
that the neutron amplitude $A$ ($A^2=\Gamma_n^0)$ would have
normal distribution with zero average and dispersion
$~~~~~D(A)=<\Gamma_n^0>$.
Nobody tested these conditions and, therefore, it is necessary to
begin analysis of the neutron width distributions just from their
obligatory test.

Again, it is postulated, but not tested that a set of the
experimental reduced neutron widths corresponds to the only one
possible distribution and does not correspond to superposition of
several functional dependences with unknown values $<A>$ and
$D(A)$.
Corresponding conclusion can be inexact within frameworks of
the results obtained in \cite{PEPAN-2006,YAF73,Prep196}.

Consequently, analysis of the distribution parameters of neutron
widths aimed to derive from them information on the neutron
resonance structure must test a possibility of presence of
superposition of $K$ distributions ($1 \le K \le 4$, for example).
This condition automatically transfers the task of distribution
analysis of neutron widths in category of search for badly
stipulated or, most probably, degenerated solutions.
Id est, serious analysis of $\Gamma_n^l$ practically inevitably
brings to multi-valued solution, and traditional analysis
(for example, \cite{2008De20}) gives solution with unknown
systematical error.

This conclusion was made from mathematical modeling of the problem
under consideration.
The expected  $<A>$, $D(A)$ and $K$ values for the accumulated
by now information on parameters of neutron resonances are
comparable with their random values  obtained at approximation of
relatively small sets of normally distributed amplitides with zero
average and unit dispersion.

Nevertheless, the full-scale analysis of the $\Gamma_n^l$ values
showed that there are no grounds to accept distribution \cite{PT}
in its classical form $<A>=0$ and $D(A)=1$ as the only true.
This conclusion follows \cite{IS18_D} from comparison of the most probable
number of resonances in experimental width distribution for
actinides at approximation of their distribution under condition
that the neutron amplitude can have non-zero average and non-unit
dispersion. The obtained in this way spacings between resonances
noticeably differ from their estimations obtained from cumulative
sums of resonances in function on neutron energy.
The exit from this situation can be found only by means of
obtaining of additional experimental information.
Id est, by realization of the methodically independent experiment.

As a tested hypothesis, it is necessary to measure the
total gamma-spectra
in different resonances and/or their groups.
The parameter for which is expected dependence on structure of wave
function of resonance can be ratio of intensity of
gamma-transitions to group of low-lying levels to mean intensity
of primary gamma-transitions with energy $E_\gamma \sim 0.5B_n$.
The experiment in spherical nuclei can be realized by use of
scintillation detectors, in deformed -- the use of Ge-detectors
is more worth while.

\section{Conclusion }\hspace*{16pt}

Analysis of condition and possibilities of modern methods for
determination of parameters of nuclear de-excitation process
followed by emission of both nucleon products and gamma-quanta
shows that the use of one-step reactions in region of high level
density allows one to obtain the $\rho$ and $\Gamma$ values only
with large systematical errors. Its maximal magnitude can be equal
to 500 - 1000\% \cite{W181,Ni60}.
The two- and multi-step reactions permit one to
decrease maximal systematical error to several tens percents.
But, only under condition that the inevitable model notions about
nucleus (on coexistence and interaction of fermi- and
bose-excitations and their influence on the process under study,
in the first turn) do not bring to noticeable and unknown errors in determination
of $\rho$ and $\Gamma$.

The volume and quality of the available data on the reduced
neutron widths do not allow one to get unambiguous conclusions
on the mean value, dispersion of neutron amplitudes and real
number of groups of resonances with different structure.
Besides, it is impossible to estimate degree of execution of
conditions which are necessary for truth of \cite{PT}.

\end{document}